\documentclass[%
superscriptaddress,
preprint,
 amsmath,amssymb,
]{revtex4-1}

\usepackage{graphicx}
\usepackage{dcolumn}
\usepackage{bm}

\begin{document}


\title{Spatiotemporal control of high-intensity laser pulses with a plasma lens}

\author{D. Li}
\email{dionli@umich.edu}
\affiliation{
G\'{e}rard Mourou Center for Ultrafast Optical Science, University of Michigan, Ann Arbor, Michigan 48109, USA}
\author{K.G. Miller}
\affiliation{
University of Rochester, Laboratory for Laser Energetics, Rochester, New York 14623-1299, USA}
\author{J.R. Pierce}
\affiliation{
Department of Physics and Astronomy, University of California, Los Angeles, California 90095, USA}
\author{W.B. Mori}
\affiliation{
Department of Physics and Astronomy, University of California, Los Angeles, California 90095, USA}
\author{A.G.R. Thomas}
\affiliation{
G\'{e}rard Mourou Center for Ultrafast Optical Science, University of Michigan, Ann Arbor, Michigan 48109, USA}
\author{J.P. Palastro}
\email{jpal@lle.rochester.edu}
\affiliation{
University of Rochester, Laboratory for Laser Energetics, Rochester, New York 14623-1299, USA}

\date{\today}

\begin{abstract}
Spatiotemporal control encompasses a variety of techniques for producing laser pulses with dynamic intensity peaks that move independently of the group velocity. This controlled motion of the intensity peak offers a new approach to optimizing laser-based applications and enhancing signatures of fundamental phenomena. Here, we demonstrate spatiotemporal control with a plasma optic. A chirped laser pulse focused by a plasma lens exhibits a moving focal point, or ``flying focus,'' that can travel at an arbitrary, predetermined velocity. Unlike currently used conventional or adaptive optics, a plasma lens can be located close to the interaction region and can operate at an orders of magnitude higher, near-relativistic intensity. 
\end{abstract}

\maketitle

\section{Introduction}
Spatiotemporal pulse shaping uses advanced optical techniques to construct laser pulses with dynamic and controllable properties \cite{Longhi03,Kondakci2017,SainteMarie2017,Froula2018,Kondakci2019,Li2020,Li2020a,Palastro2020,Jolly2020,Caizergues2020,Simpson2022,Besieris2022,Yessenov2022,Pierce2023,Ramsey2023,Liang23,Ambat2023,Pigeon2023}. The laser pulses created with these techniques can exhibit an intensity peak that travels faster than the speed of light \cite{SainteMarie2017,Froula2018}, accelerates \cite{Liang23,Ambat2023}, or even oscillates \cite{SainteMarie2017,Ambat2023}, all while maintaining a near-constant profile over distances far greater than a Rayleigh range. The flexibility to control the motion of the peak intensity has provided new opportunities to optimize laser-based applications and enhance signatures of fundamental phenomena \cite{Turnbull2018, PalastroIWAVs2018, Palastro2020,Caizergues2020,Howard19,DiPiazza2021,Ramsey2022,Formanek2022,Kabacinski2023,Simpson2023}. Nevertheless, many of these opportunities require intensities that would damage the conventional and adaptive optics used to structure the laser pulse.

Plasma optics, having been ionized, can withstand orders of magnitude higher intensities than conventional or adaptive optics \cite{Durfee1993,Malkin1999,Ping2000,Hubbard2002,Doumy2004,Levy2007,Katzir2009,Nakatsutsumi2010,Palastro2015,Turnbull2016,Lehmann2016,LeBlanc2017,Qu2017,Vieux2017,Kirkwood2018,Peng2021,Edwards2022,Edwards2022a}. The refractive index in a plasma depends on the electron density and the frequency of the laser pulse. As a result, spatial variation, temporal evolution, or nonlinearity in the electron density can be used to reflect \cite{Doumy2004,Levy2007}, refract \cite{Hubbard2002,Katzir2009,Palastro2015}, diffract \cite{LeBlanc2017,Kirkwood2018,Edwards2022a}, disperse, frequency convert \cite{Wilks1989,Peng2021,Sandberg2023}, or amplify laser pulses \cite{Malkin1999,Ping2000,Vieux2017}. In fact, several experiments already make routine use of plasma optics based on these processes: plasma gratings tune the symmetry of implosions at the National Ignition Facility \cite{Michel2009,Glenzer2010}; plasma waveguides extend the interaction lengths in laser wakefield accelerators \cite{Geddes2004,Miao2022}; and plasma mirrors enhance the intensity contrast in ultrashort pulse lasers \cite{Doumy2004,Levy2007}.

\begin{figure}
\includegraphics{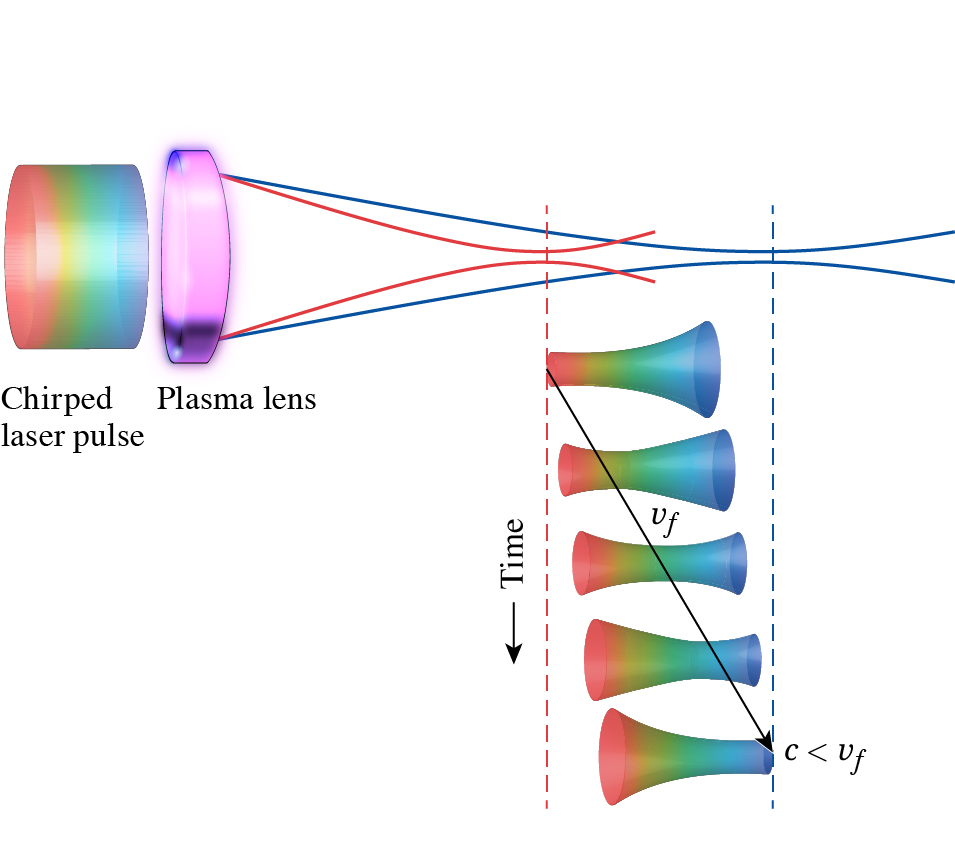}
\caption{\label{fig:f1} A chirped laser pulse focused by a plasma lens results in a moving, or ``flying,'' focus with an intensity peak that moves independently of the group velocity. The plasma lens focuses different frequencies to different longitudinal locations, while the chirp controls the arrival time of the frequencies at these locations. In this case, the velocity of the moving focus is superluminal ($v_f > c$).}
\end{figure}

Here, we demonstrate that a plasma optic can be used for the spatiotemporal control of high-intensity laser pulses. A preformed plasma channel functions as a thick, chromatic lens that focuses different frequencies in a laser pulse to different locations along the propagation axis. The chirp of the laser pulse determines the arrival time of the frequencies at these locations. The time-dependent focusing of the pulse produces a moving focal point with an arbitrary velocity that can be tuned by adjusting the chirp (Fig. \ref{fig:f1}). This configuration is a plasma-based version of the original, chromatic flying focus \cite{Froula2018}, which has been proposed for a range of experiments, including Raman amplification \cite{Turnbull2018}, photon acceleration \cite{Howard19,Franke2021}, nonlinear Thomson scattering \cite{Ramsey2022}, and vacuum birefringence \cite{Formanek2023}. However, unlike the diffractive optic used in the original flying focus, a plasma lens can be placed close to the interaction region and can operate at an orders of magnitude higher, near-relativistic intensity.

The remainder of this article begins with a model for the propagation of a laser pulse focused (or guided) by a plasma lens (Section II). The model is general enough to describe the focusing of arbitrary, space-time structured laser pulses, with or without orbital angular momentum. The model predicts that the chromatic aberration of the plasma lens and the chirp of a laser pulse can be used to produce an intensity peak with a specified, constant velocity (Section III). The plasma lens is designed to produce an extended focal region, while avoiding resonance absorption and mitigating parametric instabilities (Section IV). Particle-in-cell simulations based on this design validate the model for moderate intensities and determine the intensities and durations at which the plasma lens ceases to operate as expected (Section V). The article concludes with a summary of the results and a discussion of future prospects (Section VI). 

\section{Plasma Lens}
Consider a linearly polarized laser pulse propagating in the positive $\hat{\mathbf{z}}$ direction. The pulse is normally incident on a preformed plasma channel with an entrance and exit located at $z=0$ and $z = L_p$, respectively. The transverse electric field of the pulse can be expressed as a superposition of its frequency components:
\begin{equation}\label{eq:FT}
E(\mathbf{x},\xi) = \frac{1}{4\pi}\int e^{-i\omega\xi}\tilde{E}(\mathbf{x},\omega) d\omega +\mathrm{c.c.},
\end{equation}
where $\xi = t - z/c$ is the moving-frame coordinate.  At the entrance to the channel, the transverse profile of each component is given by
\begin{equation}\label{eq:initial}
\begin{aligned}
\tilde{E}(\mathbf{x}_{\perp},z=0,\omega) &= \sum_{q,\ell} \alpha_{q,\ell}A_I \Bigl( \frac{\sqrt{2} r}{w_I}\Bigr)^{|\ell|} L^{|\ell|}_q\Bigl(\frac{2r^2}{w^2_I}\Bigr) \\
&\text{exp}\left[ - \frac{r^2}{w_I^2} + \frac{i\omega r^2}{2cR_I} +i\ell\theta +i \phi_I \right],
\end{aligned}
\end{equation}
where $r=(x^2 + y^2)^{1/2}$, $\theta$ is the azimuth, $L_q^{|\ell|}$ is a generalized Laguerre polynomial with radial and orbital angular momentum mode numbers $q$ and $\ell$, and $\alpha_{q,\ell}$ quantifies the projection of each mode onto the initial profile. The incident amplitude $A_I$, phase $\phi_I$, spot size $w_I$, and radius of curvature $R_I$ may all depend on frequency.

Each frequency component of the pulse will refract from the plasma, advance in phase, and diffract by a different amount. This frequency-dependent evolution is described by the paraxial wave equation
\begin{equation}\label{eq:parax}
\left(2i\omega \frac{\partial}{\partial z} + c\nabla^2_{\perp}\right)\tilde{E}(\mathbf{x},\omega) = \frac{1}{c}\omega_p^2(\mathbf{x})\tilde{E}(\mathbf{x},\omega),
\end{equation}
where $\omega_p^2(\mathbf{x}) = e^2 n(\mathbf{x})/m_e\varepsilon_0$ is the square of the plasma frequency and $n(\mathbf{x})$ is the electron density. Here, the plasma response is assumed to be linear, such that $n(\mathbf{x})$ is independent of $E$ (see Section V for particle-in-cell simulations that include nonlinear effects). The plasma channel is modeled using a parabolic density profile
\begin{equation}\label{eq:profile}
n(\mathbf{x}) = n_0 + \frac{1}{2}n_2 r^2
\end{equation} 
for $0 \leq z \leq L_p$, and $n(\mathbf{x}) = 0$ otherwise. With this profile, the refractive index in the plasma, $\mu(\mathbf{x},\omega) = [1-\omega_p^2(\mathbf{x})/\omega^2]^{1/2}$ has an on-axis maximum, which bends the ``rays'' of the pulse towards the optical axis ($r=0$) like a lens. 

The transverse profile of each frequency component at any $z>0$ can be found by solving the paraxial wave equation with the initial condition in Eq. \eqref{eq:initial}. The solution is 
\begin{equation}\label{eq:evolved}
\begin{aligned}
\tilde{E}(\mathbf{x},\omega) &= \sum_{q,\ell} \alpha_{q,\ell}A(z) \Bigl[ \frac{\sqrt{2} r}{w(z)}\Bigr]^{|\ell|} L^{|\ell|}_q\Bigl[\frac{2r^2}{w^2(z)}\Bigr] \\
&\text{exp}\left[ - \frac{r^2}{w(z)^2} + \frac{i\omega r^2}{2cR(z)} +i\ell\theta +i \phi^{q,\ell}(z) \right].
\end{aligned}
\end{equation}
Within the plasma channel ($0 \leq z \leq L_p$), the frequency-dependent amplitude, phase, spot size, and radius of curvature are given by
\begin{widetext}
\begin{align}\label{eq:within1}
A(z) &= \frac{w_I}{w(z)}A_I \\
\phi^{q,\ell}(z) &= \phi_I - \frac{\omega_{p0}^2}{2c\omega}z - (2q+\ell+1) \text{atan}\left[\frac{w_I^2 Z_m}{w_m^2 R_I} + \Big(\frac{w_m^2}{w_I^2} + \frac{w_I^2 Z_m^2}{w_m^2 R_I^2 }\Big)\tan\Big(\frac{z}{Z_m}\Big)\right]   \\
w(z) &= w_I \left[1 + \frac{Z_m}{R_I}\sin\Big(\frac{2z}{Z_m}\Big) +  \Big( \frac{Z_m^2}{R_I^2} + \frac{w_m^4}{w_I^4} -1 \Big) \sin^2\Big(\frac{z}{Z_m}\Big) \right]^{1/2} \label{eq:spot} \\
R(z) &= R_I\frac{w^2(z)}{w_I^2}\left[\cos\Big(\frac{2z}{Z_m}\Big) + \frac{R_I}{2Z_m}\Big( \frac{Z_m^2}{R_I^2} + \frac{w_m^4}{w_I^4} -1 \Big)\sin\Big(\frac{2z}{Z_m}\Big)\right]^{-1}, \label{eq:curve}
\end{align}
\end{widetext}
where $w_m=(8c^2/\omega_{p2}^2)^{1/4}$ is the `matched' spot size of the plasma channel, $Z_m=\omega w_m^2 / 2c$ is the Rayleigh range associated with the matched spot size, $\omega_{p0}^2=e^2 n_0/m_e\varepsilon_0$, and $\omega_{p2}^2 = e^2 n_2/m_e\varepsilon_0$.

After the plasma channel ($z>L_p$), the frequency-dependent amplitude, phase, spot size, and radius of curvature can be expressed in terms of their values at $z=L_p$. Denoting a quantity at the exit of the plasma channel with the subscript $l$ [e.g., $A_l \equiv A(L_p)$], one finds
\begin{align}\label{eq:after1}
A(z) &= \frac{w_l}{w(z)}A_l \\
\label{eq:after2}
\phi^{q,\ell}(z) &= \phi^{q,\ell}_l- (2q+\ell+1)\text{atan}\Big(\frac{z-L_p}{Z_l}\Big) \\
\label{eq:after3}
w(z) &= w_l \left[ \Big(\frac{z - L_p + R_l}{R_l} \Big)^2 + \Big(\frac{z-L_p}{Z_l} \Big)^2  \right]^{1/2} \\
\label{eq:after4}
R(z) &= R_l \frac{w^2(z)}{w_l^2}\left[1 + \Big(1 + \frac{R_l^2}{Z_l^2}\Big)\Big(\frac{z-L_p}{R_l}\Big)\right]^{-1}, 
\end{align}
where $Z_l = \omega w_l^2 / 2c$. Equations \eqref{eq:after1}--\eqref{eq:after4} describe the focusing (or defocusing) of an arbitrary Laguerre-Gaussian mode by a lens with a focal length 
\begin{equation}\label{eq:fl}
f_l = -\frac{R_l}{1+R_l^2/Z_l^2}
\end{equation}
to a minimum spot size
\begin{equation}\label{eq:wmin}
w_f = \frac{2cf_l}{\omega w_l}\bigg(\frac{2}{1+\sqrt{1-4f_l^2/Z_l^2}}\bigg)
\end{equation}
in the plane $z=L_p + f_l$. Thus, a plasma channel can be used as a lens.

For nearly all parameters of interest, $|R_I| \gg Z_I \gg Z_m$ and $Z_l \gg |R_l|$ (see Appendix A). Under these conditions, the focal length and focused spot size of the plasma lens reduce to the relatively simple expressions 
\begin{align}\label{eq:fl}
f_l(\omega) &\approx \frac{\omega w_m^2}{2c} \cot\Big(\frac{2cL_p}{\omega w_m^2}\Big) \\ \label{eq:wl}
w_f(\omega) &\approx \frac{w_m^2}{w_I}\left| \csc\Big(\frac{2cL_p}{\omega w_m^2}\Big) \right|,
\end{align}
for $j\pi \leq L_p/Z_m \leq (j+\tfrac{1}{2})\pi$ and $j$ an integer, which ensures $f_l > 0$. If in addition $L_p \ll Z_m$, the laser pulse does not appreciably diffract or refract within the plasma lens, $w_l \approx w_I$, and Eqs. \eqref{eq:fl} and \eqref{eq:wl} further reduce to  $f_l(\omega) \approx Z_m^2/L_p$ and $w _f(\omega) \approx 2cf_l/\omega w_l$. This is the ``thin'' lens limit.  Otherwise, the plasma lens is considered ``thick.'' Regardless of the thickness, the focal length of the plasma lens depends on frequency, i.e., the lens is chromatic.

\section{Plasma Lens Flying Focus}
The plasma lens focuses each frequency component of the laser pulse to a different longitudinal location $z_f(\omega) = L_p + f_l(\omega)$. This produces an extended focal range with a length determined by the minimum and maximum frequencies: 
\begin{equation}\label{eq:Lf}
L_f = f_l(\omega_{\text{max}}) - f_l(\omega_{\text{min}}).
\end{equation}
Each frequency arrives at its focal location at a different time $t_f(\omega)$. The focal time is the sum of two contributions: the relative timing of each frequency within the laser pulse at the exit of the plasma lens, i.e., $\partial_{\omega}\phi^{q,\ell}_l(\omega)$, and the time it takes each frequency to travel a distance $f_l(\omega)$. In total,
\begin{equation}\label{eq:tf}
t_f(\omega) = \frac{1}{c}f_l(\omega) + \partial_{\omega}\phi^{q,\ell}_l(\omega).
\end{equation}
The frequency-dependent focal location and time results in a moving focal point with a velocity
\begin{equation}\label{eq:vf}
v_f(\omega) = \frac{dz_f}{d\omega}\Big(\frac{dt_f}{d\omega}
\Big)^{-1} = c\left[1 + c\Big(\frac{\partial \omega}{\partial f_l}\Big)\Big(\frac{\partial^2\phi^{q,\ell}_l}{\partial \omega^2}\Big)\right]^{-1}.
\end{equation}
The focal velocity can be tuned by adjusting the properties of the plasma lens, the mode numbers, or the initial spectral phase of the laser pulse $\phi_I(\omega)$. 

Up to this point, the analysis has considered a laser pulse with an arbitrary frequency spectrum. For the remainder, it is convenient to consider a laser pulse characterized by a ``central'' frequency $\omega_0$ and spectral width $\tau^{-1} \ll \omega_0$. When the conditions $|R_I| \gg Z_I \gg Z_m$ and $Z_l \gg |R_l|$ are satisfied (see Appendix A), the spectral phase at the exit of the plasma lens is independent of $q$ and $\ell$:
\begin{equation}\label{eq:philsimple}
\phi^{q,\ell}_l(\omega) \approx \phi_I(\omega) - \frac{\omega_{p0}^2}{2c\omega}L_p \equiv \phi_l(\omega),
\end{equation}
where constant phase terms have been dropped. Because the focal velocity depends on $\partial_{\omega}^2 \phi_l$, an initial second-order spectral phase 
\begin{equation}\label{eq:spphase}
\phi_I(\omega) = \frac{1}{2}\phi_2(\omega-\omega_0)^2
\end{equation}
is the simplest phase that provides control over the velocity. This is equivalent to chirping the laser pulse. More specifically, one can write $\phi_2 = \hat{\eta} \tau^2 /2$, where the chirp parameter $\hat{\eta}$ quantifies the temporal elongation of the pulse. 

\begin{figure*}
\includegraphics[scale=0.85]{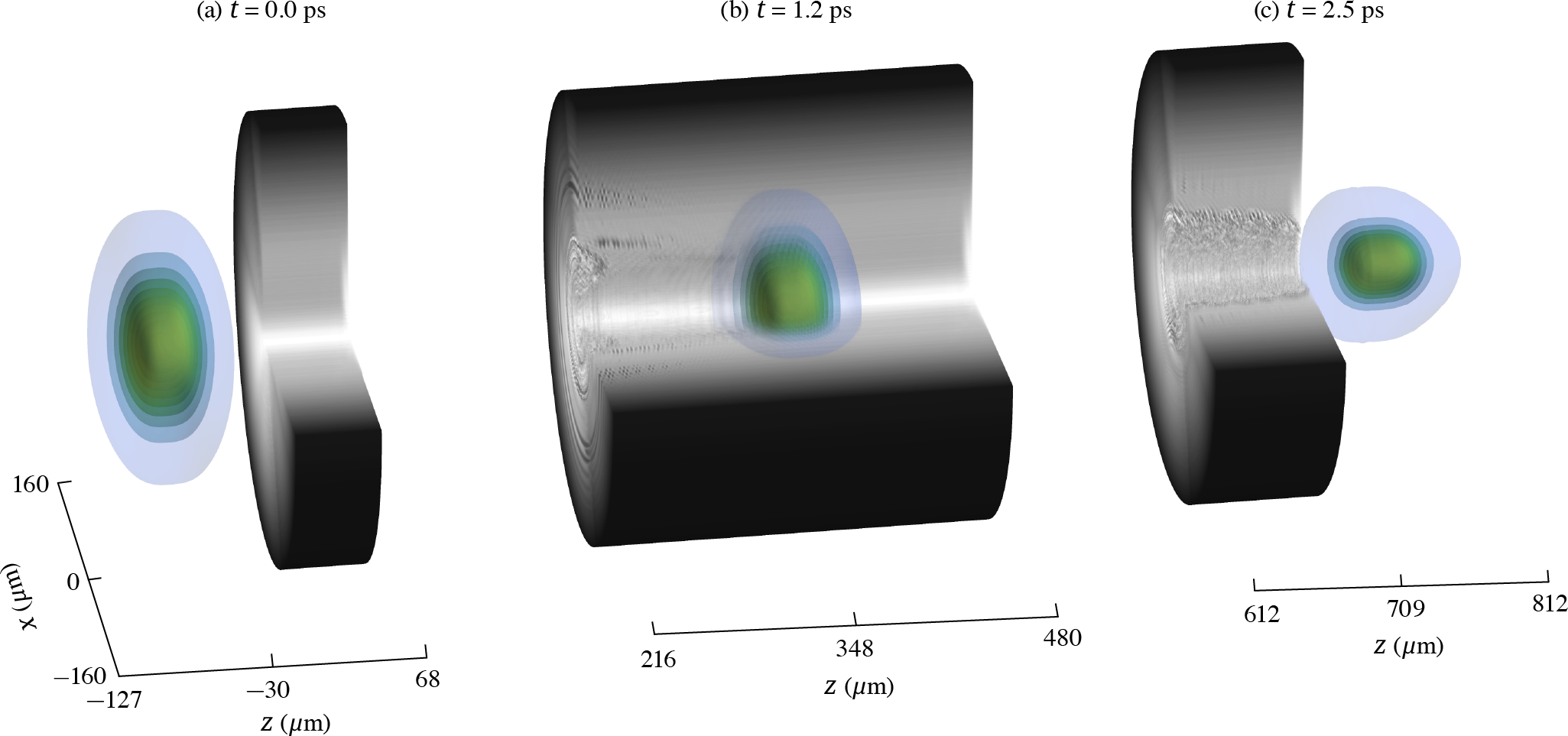}
\caption{\label{fig:f2} Density contours of the plasma lens (grey scale) and intensity contours (color scale) of a laser pulse traversing the plasma lens. For this design, the plasma lens is ``thick,'' and the spot size of the laser pulse decreases appreciably from the entrance to the exit of the lens (left to right). The incident laser pulse has a normalized amplitude $a_I = 0.5$ and chirp parameter $\eta = -5$. All other parameters can be found in Table I. }
\end{figure*}

The plasma contribution to the spectral phase [Eq. \eqref{eq:philsimple}] can be compensated by the initial spectral phase $\phi_I$. To second order in $\omega - \omega_0$,
\begin{equation} \label{eq:philcomp}
\phi_l(\omega) \approx \frac{1}{4}\left(\hat{\eta} -\frac{2\omega_{p0}^2L_p}{c\omega_0^3\tau^2}\right)\tau^2(\omega-\omega_0)^2,
\end{equation}
where constant and linear phase terms have been dropped (the latter would only result in an overall temporal delay). Thus, setting the chirp parameter $\hat{\eta} = \eta + 2\omega_{p0}^2 L_p/c\omega_0^3\tau^2$ compensates the second-order phase acquired in the plasma lens and results in a spectral phase determined solely by $\eta$. Higher-order phase could also be compensated by introducing higher-order terms in $\phi_I$. However, for the cases considered here, the second-order phase is already small: $2\omega_{p0}^2 L_p/c\tau^2\omega_0^3 \ll 1$.

With the compensated spectral phase [Eq. \eqref{eq:philcomp}], the chromatic focusing of the chirped pulse results in the focal velocity
\begin{equation}\label{eq:vfred}
\frac{v_f(\omega)}{c} = \left[1 + \frac{\eta(\frac{c\tau}{w_m})^2\frac{Z_m}{L_p}\sin^2(\frac{L_p}{Z_m})}{1+\frac{Z_m}{2L_p}\sin(\frac{2L_p}{Z_m})}\right]^{-1}.
\end{equation}
The focal velocity depends on frequency through the matched Rayleigh range $Z_m = \omega w_m^2 / 2c$. To first approximation, this dependence can be ignored, and $\omega$ can be replaced by $\omega_0$ because $\omega_0 \tau \ll 1$. As a result, the focal velocity is nearly constant throughout the focal range $L_f$.

For large values of the chirp parameter ($\eta \gg 1$), the stationary phase approximation can be used in Eq. \eqref{eq:FT} to find the electric field of the laser pulse in the time domain:
\begin{equation}\label{eq:SPA}
E(\mathbf{x},\xi) \approx \frac{\text{sgn}(\eta)}{(4\pi|\eta|\tau^2)^{1/2}}\tilde{E}(\mathbf{x},\omega_s)  e^{-i(\omega_s\xi - \frac{\pi}{4})} + \mathrm{c.c.},
\end{equation}
where $\omega_s = \omega_0 + 2\xi/\eta\tau^2$. The time-domain envelope of the laser pulse $\tilde{E}(\mathbf{x},\omega_s)$ is given by Eq. \eqref{eq:initial} with $\omega$ replaced by $\omega_s$. Using Eqs. \eqref{eq:within1}, \eqref{eq:after1}, and the replacement $\omega \rightarrow\omega_s$ in Eq. \eqref{eq:after3} yields the temporal profile of the moving focus within the focal range: 
\begin{equation}\label{eq:IF}
I_f \propto \frac{w_I^2A_I^2}{w^2(z)} \approx \frac{w_I^2A_I^2}{w_f^2} \left[ 1 + \Big(\frac{t - z/v_f - \zeta_0}{\tau_f} \Big)^2  \right]^{-1},
\end{equation}
where 
\begin{equation}\label{eq:taue}
\tau_f = \left|\frac{c-v_f}{cv_f}\right|\frac{\omega_0w_f^2}{2c}
\end{equation} 
is the duration of the intensity peak
and $\zeta_0 = (v_f-c)[L_p + f_l(\omega_0)]/cv_f$ is the focal time with respect to the coordinate $\zeta = t - z/v_f$ for $\omega=\omega_0$. The maximum value of the moving intensity peak will be approximately constant within the focal range if the spectral amplitude $A_I(
\omega)$ is constant for $\omega_{\mathrm{min}} \leq \omega \leq \omega_{\mathrm{max}}$. 

\section{Design Considerations}
A distinguishing property of a flying focus is that the peak intensity can maintained over a distance greater than the Rayleigh range of the focused spot. To take advantage of this property and to ensure that the focal trajectory can be clearly identified, the plasma lens should be designed so that 
\begin{eqnarray}\label{eq:cond1}
L_f > \frac{\omega_0w_f^2}{2c}.
\end{eqnarray}
In addition, the parameters for the plasma lens should be chosen to avoid resonance absorption and mitigate parametric instabilities. This can be accomplished by keeping the maximum density experienced by the pulse $n_{
\text{max}} \sim n_2w_I^2$ well below the critical density $n_{cr} = m_e\varepsilon_0\omega_0^2/e^2$. 
Equation \eqref{eq:cond1} can be re-expressed in terms of these densities and the length of the plasma lens as follows:
\begin{eqnarray}\label{eq:cond2}
\left(\frac{\Delta \omega L_p }{2c}\right)\left(\frac{n_{\text{max}}}{n_{cr}}\right) > 1,
\end{eqnarray}
where $\Delta \omega \equiv \omega_{\text{max}} - \omega_{\text{min}}$ is the total bandwidth of the laser pulse and $L_f \approx \Delta \omega \partial_{\omega} f_l$ has been used. Preventing instabilities like two-plasmon decay and absolute stimulated Raman scattering  requires $n_{\text{max}}/n_{cr} \lesssim 1/4$. Thus, for a fixed bandwidth $\Delta \omega$, the length of the focal region relative to the Rayleigh range is determined solely by $L_p$. Note, however, that (1) $L_p$ can only be increased up to $\pi Z_m/2$ before the plasma lens becomes a defocusing lens [see Eq. \eqref{eq:fl}], and (2) a longer plasma lens may exacerbate instabilities like stimulated Raman forward scattering.

With these considerations in mind, a plasma lens was designed to produce flying foci over a range $L_f = \omega_0 w_f^2 / c $. The parameters, displayed in Table I, were motivated by commercially available Ti:sapphire laser systems and experimentally demonstrated plasma channels \cite{Durfee1993,Durfee1995}. The rightmost column of the table provides the parameters in normalized units to facilitate scaling the results presented here to other laser wavelengths or plasma densities. For this design, $L_p \approx Z_m$. As a result, the plasma lens is ``thick,'' and the spot size evolves significantly within the lens (Fig. \ref{fig:f2}).

\begin{table}[b]
\caption{\label{tab:table1}
Laser pulse and plasma lens parameters used in the simulations. In the rightmost column, space, time, and density are normalized by $c/\omega_{0}$, $1/\omega_{0}$, and $n_{cr} = m_e\varepsilon_0\omega_0^2/e^2$. The vacuum wavelength $\lambda_0 = 2\pi c/\omega_0$. The chirp parameter $\eta$ was varied to change the focal velocity.}
\begin{ruledtabular}
\begin{tabular}{ccc}
Pulse parameters & Value & Normalized \\
\hline
$\lambda_0$ & 800 nm & 2$\pi$ \\
$\omega_0$ & $2.4\times10^{15}$ rad/s & 1 \\
$\Delta \omega$ (FWHM) & $1.4\times10^{14}$ rad/s & 0.061 \\
$\tau$ & 21 fs & 51 \\
$R_I$ & $\infty$ & $\infty$ \\
$w_I$ & $60$ $\mu$m & 470 \\

\hline
Plasma lens parameters & Value & Normalized \\
\hline
$n_0$ & $1\times10^{18} \text{cm}^{-3}$ & 5.7$\times10^{-4}$ \\
$w_m$ & $12.5$ $\mu$m & 98 \\
$L_p$ & $0.61$ mm & 4.8$\times10^3$  \\
$f_l(\omega_0)$ & $0.4$ mm  & 3.1$\times10^3$ \\
$w_f(\omega_0)$ & 3.1 $\mu$m & $24$ \\
$L_f$ & $77$ $\mu$m & 604 \\
$f/\#$ & 6.1 & 6.1 \\
\end{tabular}
\end{ruledtabular}
\end{table}

\begin{figure}
\includegraphics{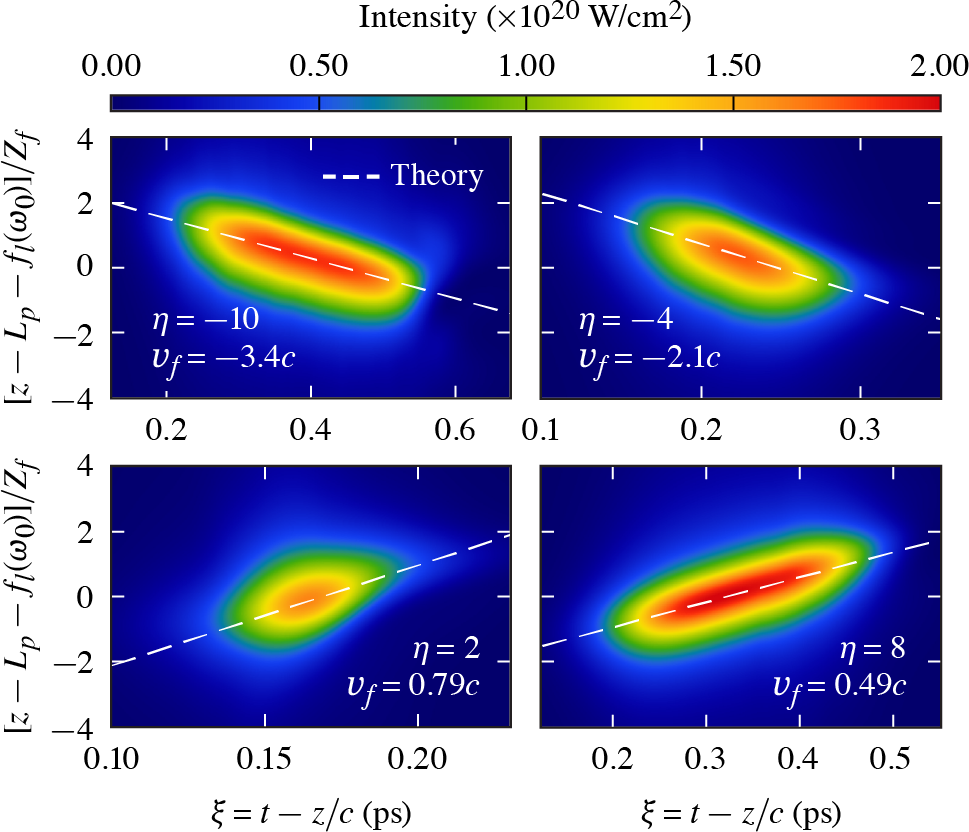}
\caption{\label{fig:f3} Spatiotemporal profiles of the moving focus within the focal range for different focal velocities $v_f$ (or chirps $\eta$) at $r=0$. The intensity peak has a near-constant velocity and a maximum intensity that is consistent with the theory. In each case, the incident laser pulse had a normalized amplitude $a_I = 0.5$. The distance from the nominal focus $f_l(\omega_0)$ is normalized to the Rayleigh range of the focused spot at the central frequency $Z_f \equiv cw_f^2/2\omega_0$.}
\end{figure}

\section{Simulation Results}
The model presented in the previous sections illustrates the salient phenomena underlying a plasma-lens-based flying focus. Nevertheless, the model neglects non-paraxial and nonlinear propagation effects, such as stimulated Raman scattering, self-focusing, ponderomotively driven density modifications, and the increase in the effective electron mass due to relativistic motion. This section presents the results of quasi-3D particle-in-cell (PIC) simulations that include these effects (see Appendix B for details). 

For incident intensities up to $I_I = 5.4\times10^{17} \; \text{W/cm}^2$, the simulations demonstrate that the plasma-lens-based flying focus works as designed, validating the model. For incident intensities $I_I \gtrsim 2\times10^{18} \; \text{W/cm}^2$, the simulations show that nonlinear effects disrupt the focusing of the plasma lens and formation of a flying focus. Notably, these intensities correspond to normalized amplitudes $a_I \equiv eE_I/m_e c \omega_0$ equal to  0.5 and 1.0, respectively, where $E_I = \text{max}[E(\mathbf{x}_{\perp},z=0,\xi)]$, which straddle the transition from non-relativistic to relativistic electron motion. 

The incident laser pulse was initialized in the frequency domain as in Eq. \eqref{eq:initial} with a transverse Gaussian profile ($q=\ell=0$). The initial spectral amplitude had the super-Gaussian profile 
\begin{eqnarray}\label{eq:initprofSG}
A_I(\omega) = \exp\left\{- [\tfrac{1}{2}\tau(\omega - \omega_0)]^4 \right\}, 
\end{eqnarray}
with a corresponding full width at half maximum $\Delta \omega = (4/\tau)[\text{ln}(2)/2]^{1/4}$. The relatively flat spectral amplitude was chosen to ensure a near-constant peak intensity within the focal region. The initial spectral phase was specified as in Eq. \eqref{eq:spphase} with $\phi_2 = \eta \tau^2 /2$. The on-axis plasma density $n_0$ was low enough that the plasma contributed a negligible second-order phase. All other parameters can be found in Table I.

Figure \ref{fig:f3} demonstrates that a plasma lens can produce a moving intensity peak with a predesigned velocity $v_f$. The panels show the on-axis ($r=0$) spatiotemporal profile of the intensity peak as it traverses the focal range for four different chirp values. In each case, the moving focus has a near-constant velocity as predicted by Eq. \eqref{eq:vfred} (white dashed lines). Note that with respect to the moving-frame coordinate $\xi = t - z/c$, the intensity peak follows the trajectory $z = cv_f\xi/(c-v_f)$. 

\begin{figure}
\includegraphics{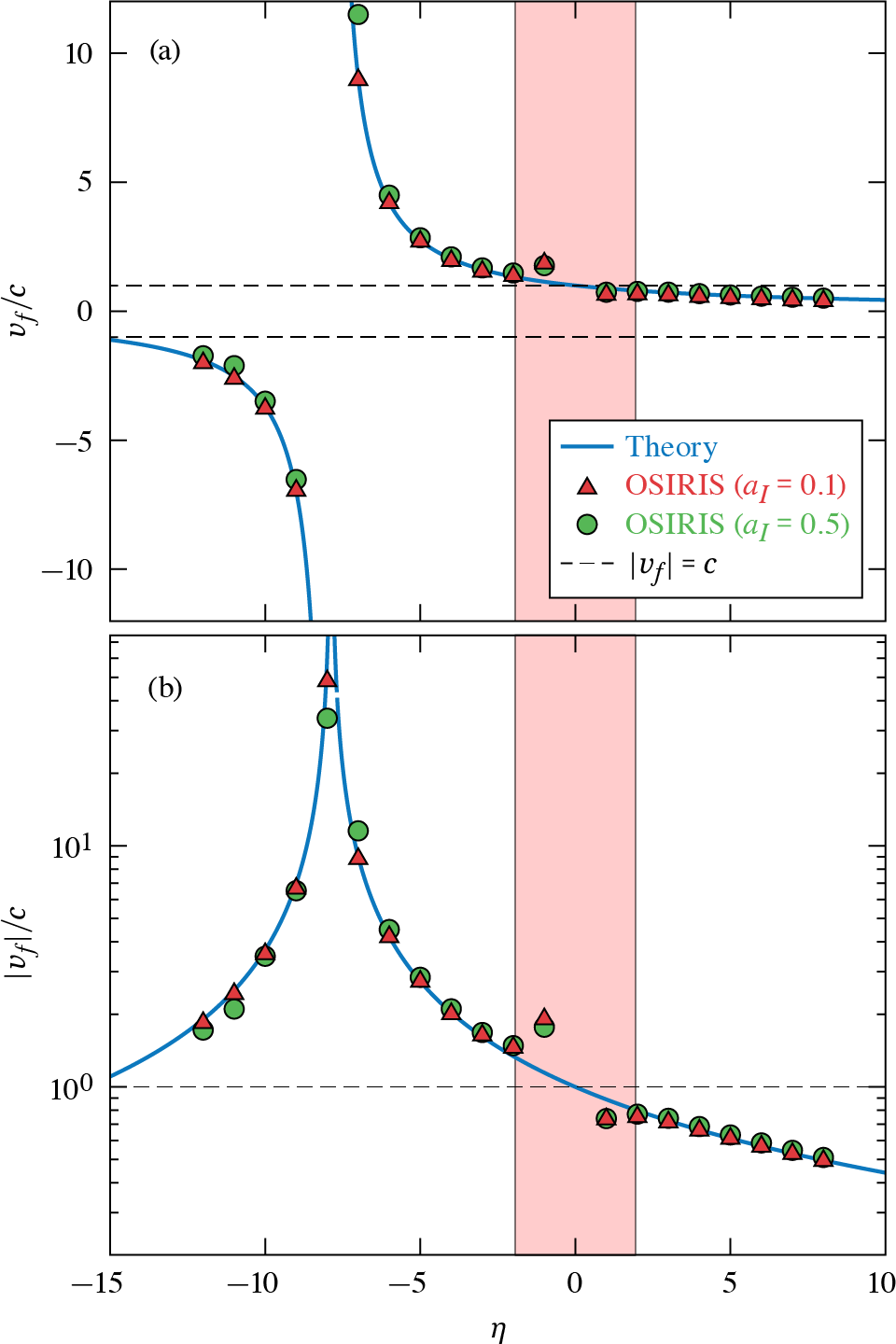}
\caption{\label{fig:f4} Comparison of the focal velocities predicted by the ``thick'' lens theory [Eq. \eqref{eq:vfred}] and the PIC simulations as a function of the chirp parameter $\eta$. The theory and simulations are in excellent agreement when the effective duration $\tau_f$ is larger than the stretched pulse duration (outside the red shaded area).}
\end{figure}

Figure \ref{fig:f4} compares the focal velocities calculated from the theory and simulations for all simulated chirp values. For focal velocities ranging from negative to positive superluminal, the two are in excellent agreement. Slight discrepancies can be observed for small values of the chirp parameter ($|\eta|\le2$). At these chirps, the effective duration of the moving focus $\tau_f$ [Eq.\eqref{eq:taue}] becomes larger than the duration of the laser pulse. In addition, the temporal profile of the laser pulse begins to approach the transform-limited profile, which no longer resembles the spectral amplitude (i.e., the stationary phase approximation breaks down). The combination of these effects makes it difficult to discern the intensity peak of the moving focus from the inherent intensity peak of the temporal profile. 

In Fig. \ref{fig:f3}, the incident laser pulses had an amplitude $a_I = 0.5$, corresponding to an intensity of $I_I = 5.4\times10^{17} \; \text{W/cm}^2$. For the larger chirp values, the peak intensity at focus is consistent with Eqs. \eqref{eq:within1} and \eqref{eq:after1}: $I_f = (w_I/w_f)^2I_I = 2\times10^{20} \, \text{W/cm}^2$. The slightly lower intensity when $\eta = 2$ results from modifications to the temporal profile not captured within the stationary phase approximation as discussed above (i.e., the temporal profile is approaching its transform limit). 

As the incident amplitude of the laser pulse is pushed beyond $a_I = 0.5$, nonlinear propagation within the plasma lens and the nonlinear plasma response begin to disrupt the formation of a flying focus. Figure 5 shows the on-axis intensity profile at $z = L_p + f_l(\omega_0)$ normalized to the maximum of the incident intensity for $a_I =$ 0.1, 0.5, and 1.0, and three stretched pulse durations (chirp values). For the shorter pulse durations ($\eta = -4$ and $-8$), the $a_I = 0.1$ and $a_I = 0.5$ profiles are nearly identical, indicating that nonlinear effects have not modified the plasma lens focusing. For the longest pulse duration ($\eta = -12$), the profile of the $a_I = 0.1$ pulse remains unmodified, while the back half of the $a_I = 0.5$ pulse has become distorted. When $a_I = 1$, this distortion occurs earlier within the pulse and is already apparent at the intermediate duration $\eta = -8$. At the longest pulse duration ($\eta = -12$), the entire back half of the pulse has deteriorated. 

Mitigating these nonlinear modifications to the laser pulse requires either lowering the incident amplitude or shortening the stretched pulse duration. To shorten the duration for a fixed focal spot size (f-number) and focal velocity [Eqs. \eqref{eq:wl} and  \eqref{eq:vfred}], one can use the scalings $\tau \rightarrow \chi \tau$, $\Delta \omega \rightarrow \chi^{-1} \Delta \omega$ , $w_m \rightarrow \chi w_m$, $L_p \rightarrow \chi^2 L_p$, $w_I \rightarrow \chi^2 w_I$, and $n_{\text{max}} \rightarrow n_{\text{max}}$. With these scalings, the right-hand side of Eq. \eqref{eq:cond2} $ \rightarrow \chi (\Delta \omega L_p n_{\text{max}})/(2cn_{cr})$. This scaling shows that decreasing the duration ($\chi < 1$) makes it more difficult to satisfy Eq. \eqref{eq:cond2}. As a result, lowering the amplitude may be the preferable option for mitigating nonlinear modifications to the pulse.

\begin{figure}
\includegraphics{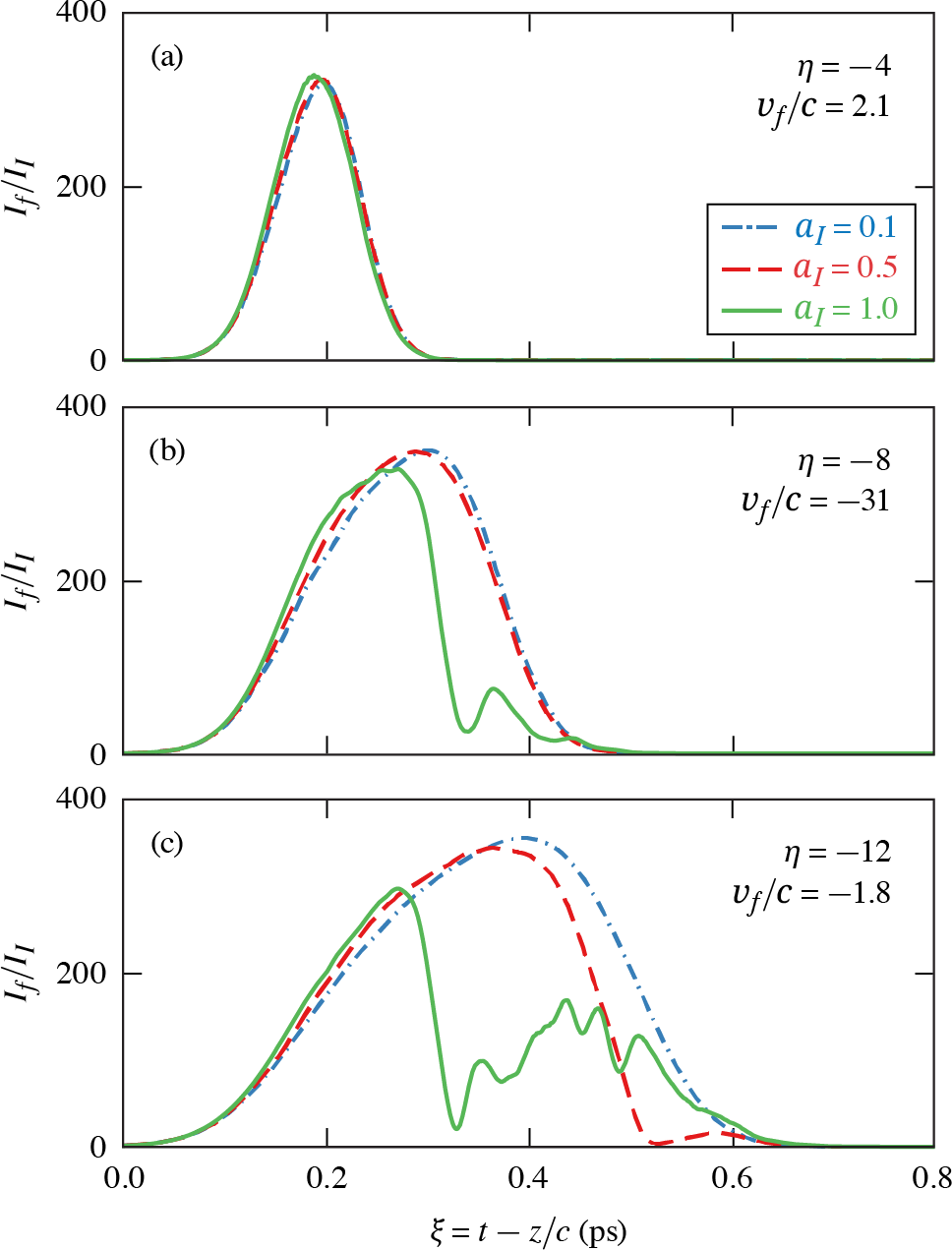}
\caption{\label{fig:f5} Temporal profiles of the on-axis intensity at the nominal focal point $z = L_p + f_l(\omega_0)$ for three different focal velocities (or chirps $\eta)$ and incident amplitudes $a_I$. In all cases, the intensity is normalized to the incident intensity $I_I$. Pulses with longer durations and larger amplitudes are more susceptible to modifications due to nonlinear propagation and plasma evolution.}
\end{figure}

\section{Conclusions and Prospects}
Plasma optics allow for spatiotemporal control at orders of magnitude higher intensities than conventional optics. In the specific case considered here, a chirped laser pulse focused by a plasma lens exhibits a dynamic or ``flying'' focus that moves independently of the group or phase velocities. By adjusting the chirp, the velocity of the moving focus can be varied from sub to superluminal in either the forward or backward directions. The plasma lens can be created using the same experimental techniques used to produce plasma channels and operates as designed up to near-relavistic intensities ($I \approx 6\times10^{17} \, \mathrm{W/cm^2}$).

The parameters of the plasma lens considered were motivated by experimentally demonstrated plasma channels \cite{Durfee1993,Durfee1995}. With these parameters, the incident intensity $I_I = 5.4\times10^{17} \, \mathrm{W/cm^2}$ resulted in a focused intensity of $I_f = 2.0\times10^{20} \, \mathrm{W/cm^2}$. The focused intensity can be increased or decreased by scaling the parameters of the plasma lens. For a fixed incident intensity and focused spot size (or f-number), the scalings $w_I \rightarrow \chi w_I$, $w_m \rightarrow \chi^{1/2} w_m$, $L_p \rightarrow \chi L_p$, and  $f_l \rightarrow \chi f_l$ result in a focused intensity $I_f \rightarrow \chi^2 I_f$. As an example, achieving a plasma-lens-based flying focus with a world-record intensity of $I_f = 1\times10^{23} \, \mathrm{W/cm^2}$ \cite{Yoon2021} would require a plasma lens with a maximum radius of $r_{\mathrm{max}} \sim w_I \sim 1.2 \, \mathrm{mm}$ and a length of $L_p \sim 1.2$ cm.

As an alternative to plasma channels, a plasma lens can also be created by using the interference of two laser beams to structure the plasma density through ionization or ponderomotive forces \cite{Edwards2022}. These holographic plasma lenses operate like a diffractive lens and are inherently chromatic. A chirped laser pulse focused by such a plasma lens would also produce a flying focus. Future work will consider this alternative and explore other plasma configurations that can controllably modify the space-time structure of laser pulses. For instance, a plasma-based version of the ultrafast flying focus \cite{Palastro2020} or the flying focus X \cite{Simpson2022} could allow for arbitrary focal velocities and an ultrashort-duration intensity peak that travels distances much greater than a Rayleigh range.

\begin{acknowledgments}
The authors would like to thank J. Vieira, T.T. Simpson, D. Ramsey, and D.H. Froula for productive discussions. The work of DL and AGRT is supported by the National Science Foundation under Award Number NSF-2108075 and the Office of Fusion Energy Sciences under Award Number DE-SC0022109. The work of JRP and WBM is supported by the LLE subcontract SUB00000211/GR531765, the National Science Foundation under Award Number NSF-2108970, and the Office of Fusion Energy Sciences under Award Number DE-NA0004131. The work of KGM and JPP is supported by the Office of Fusion Energy Sciences under Award Number DE-SC00215057, the Department of Energy National Nuclear Security Administration under Award Number DE-NA0004144, the University of Rochester, and the New York State Energy Research and Development Authority.
\end{acknowledgments}

\appendix

\section{Operating Regime}

This appendix justifies the conditions $|R_I| \gg Z_I \gg Z_m$ and $Z_l \gg |R_l|$, which were used to simplify the equations derived in Sections II and III. In any situation of interest, a plasma lens would be located far from an initial optical assembly. Suppose then that the laser pulse incident on the plasma lens was originally collimated and subsequently focused by a conventional lens with a focal length $f_I \approx |R_I|$ located at $z = -z_0 <0$. At the entrance to the plasma lens, the spot size would be
\begin{equation}\label{eq:conspot}
w_I \approx w_0 \left[ \Big(\frac{\Delta}{f_I} \Big)^2 + \Big(\frac{w_D}{w_0} \Big)^2 \right]^{1/2},
\end{equation}
where $w_0$ is the spot size incident on the conventional lens, $\Delta \equiv |f_I -z_0| \ll f_I$ is the distance between the conventional focus and the entrance to the plasma lens, and $w_D = 2 c f_I/ \omega w_0$ is the diffraction-limited spot size. If the entrance to the plasma lens is located within the confocal region of the conventional lens, i.e., $\Delta \leq \omega w_D^2/2c$, then $w_I \approx w_D$. As a result, $|R_I|/Z_I \approx f_I/Z_I \approx w_0/w_D$. For the lenses of interest $w_0 \gg w_D$, such that $|R_I| \gg Z_I$. If the plasma lens is outside of the confocal region,  i.e., $\Delta > \omega w_D^2/2c$, then $w_I \approx w_0 \Delta/f_I$. In this case,  $|R_I|/Z_I \approx f_I/Z_I \approx (w_D/w_0)(f_I/\Delta)^2$, which easily satisfies the first condition when the plasma lens is far from the conventional lens: $\Delta \ll (w_D/w_0)^{1/2}f_I$.

The condition $Z_I \gg Z_m$ is equivalent to $w_I^2/w_m^2 \gg1$. When $|R_I|/Z_I \gg 1$, the spot size of the laser pulse can oscillate between $w_I$ and $w_m^2/w_I$, depending on the length of the plasma lens [Eq. \eqref{eq:spot}]. A ratio $w_I/w_m > 1$ ensures that the laser pulse exits the plasma lens with a smaller spot than it entered with. Perhaps more importantly, the radius of curvature can oscillate in the plasma lens [Eq. \eqref{eq:curve}]. These oscillations can result in unintended defocusing of the laser pulse. When $R_I < 0$ (e.g., due to preliminary focusing by a conventional lens), $w_I > w_m$ ensures that the laser pulse accumulates a stronger focusing phase over the initial length of the plasma lens. This initial length $L_i$ and the ultimate strength of the focusing phase reach their maximum in the limit $w_I^2/w_m^2 \rightarrow \infty$. More specifically, $L_i \rightarrow \pi Z_m /2$ and $R \rightarrow 0$ from below. Thus, operating in a regime where $w_I^2/w_m^2 \gg1$ provides a larger range of available focusing powers and a larger margin on the length of the plasma before one needs to worry about a sign reversal in the curvature phase. Finally, for any focusing plasma lens of use, the focused spot size $w_f$ will be much smaller than the spot size at the exit of the lens $w_l$, such that $Z_l / |R_l| = \omega w_l^2 / 2c|R_l| \approx w_l / w_f \gg 1$. 
\vspace{5pt}

\section{Simulation Details}
All of the simulations presented in this work were performed using \textsc{osiris} with the moving-window and quasi-3D capabilities \cite{Fonseca2002,Fonseca2013,Davidson2015}.
The transverse domain consisted of two azimuthal modes and 600 cells over $240 \; \mu$m in the radial direction. The longitudinal domain was scaled with the chirp to contain the stretched duration of the laser pulse. The minimum and maximum number of longitudinal cells were 3200 and 9600, corresponding to lengths of 80 and 240 $\mu$m. The longitudinal resolution was fixed at $\Delta\xi\sim$25 nm to maintain 30 grid points per the shortest wavelength in the pulse. 

The electromagnetic fields and particle motion were evolved with the dual solver, which ensures accurate dispersion of the waves and eliminates time-staggering errors in the Lorentz force \cite{Li2021}. To accomplish this, the dual solver employed finite-difference operators with 16 coefficients. A 0.083 fs time step was set to satisfy the Courant condition, and the simulation duration was $\sim$6 ps. Open boundary conditions were applied for both the fields and particles. In the region occupied by the plasma lens, the ions were fixed, and 32 particles per cell were used for the electrons. The plasma lens had $10\,\mu$m density ramps at its entrance and exit.

\bibliography{bib}

\end{document}